\documentclass{article}

\usepackage{graphicx}
\usepackage{dcolumn}
\usepackage{amsmath}
\usepackage{amssymb}
\usepackage{array}

\newcommand{\real}{{\mathbb R}}
\newcommand{\ham}{{\mathbb H}}
\newcommand{\alg}{{\mathbb A}}
\newcommand{\mat}[1]{\left(
\begin{array}{ccc}
#1
\end{array}
\right)}

\begin{document}

\small

\title{Born-Infeld Lagrangian using Cayley-Dickson algebras}

\author{S. Kuwata}
\date{Faculty of Information Sciences, Hiroshima City University\\
Asaminami-ku, Hiroshima 731-3194, Japan}


\maketitle

\begin{abstract}

We rewrite the Born-Infeld Lagrangian, which is originally given by the
determinant of a $4 \times 4$ matrix composed of the metric tensor $g$ and the
field strength tensor $F$, using the determinant of a $(4 \cdot 2^n)
\times (4 \cdot 2^n)$ matrix $H_{4 \cdot 2^{n}}$. If the elements of
$H_{4 \cdot 2^{n}}$ are given by the linear combination of $g$ and $F$, it is
found, based on the representation matrix for the multiplication operator of
the Cayley-Dickson algebras, that
$H_{4
\cdot 2^{n}}$ is distinguished by a single parameter, where distinguished
matrices are not similar matrices.
%
We also give a reasonable
condition to fix the parameter.

\end{abstract}

\section{Introduction}

The Born-Infeld Lagrangian~\cite{Born}, which was originally introduced as a
possible short-distance modification of the electromagnetic interaction, is
written using the determinant of the $4 \times 4$ matrix which is composed of
the metric tensor $g$ and the field strength tensor $F$.
The Born-Infeld action appears as the
effective low energy action for the gauge field
on D-branes in the string theory~\cite{string}, where non-commutative geometry has
been used in formulating open strings~\cite{witten}.
In the context of the string theory, non-commutative Yang-Mills theory on a torus
has been extracted~\cite{connes,douglas}.

Recently, it has been pointed out by Schuller~\cite{schuller} that
non-commutative spacetime is directly derived from the Born-Infeld Lagrangian by
rewriting it using the
$8 \times 8$ matrix whose 64 elements are given by the commutator of
8-dimensional phase space variables.
The $8 \times 8$ matrix is chosen as antisymmetric, so that the elements can
be represented as the commutator of some variables (in this case, phase space
variables). Although the finding that the Born-Infeld dynamics is closely related to
the non-commutative spacetime is intriguing in itself,
this raises a problem of whether the
$8 \times 8$ matrix are uniquely determined (to be antisymmetric) or
not, excluding equivalent matrices obtained through a similarity
transformation.
If not, some condition is required for the $8 \times 8$ matrix to be
equivalently represented.

On the other hand, the original $4 \times 4$ matrix in the (Euclideanized)
Born-Infeld Lagrangian can be written using the representation matrix
for some quaternion, which is the subset of the octonion (both are members
of the Cayley-Dickson algebras). This
implies that the Cayley-Dickson algebras play a useful role in solving the
above problem. Even if we restrict the
$8 \times 8$ matrix to such that its four blocks of $4 \times 4$ matrices
are given by the linear combination of $g$
and $F$ (this means that the four $4 \times 4$ matrices are rank-2 tensors of
the same type of $g$ and $F$), it will be found that the answer is still
negative; the obtained
$8 \times 8$ matrix is distinguished by a single parameter, where
distinguished matrices cannot be obtained from each other through a
similarity transformation. Furthermore, it is tempting to obtain a desired
square matrix whose dimension is 16, 32, and more.
By considering the higher dimensional matrix, there is a possibility that the more
new interpretation may be realized for the Born-Infeld dynamics.

The aim of this article is that we first rewrite, based on the
representation matrix for the multiplication operator of the Cayley-Dickson
algebras, the Born-Infeld Lagrangian using the determinant of the $(4 \cdot
2^n) \times (4
\cdot 2^n)$ matrix whose
$4^{n}$ blocks of
$4
\times 4$ matrix are given by the linear combination of $g$ and $F$.
Despite the number of $n \; ( \geq 1)$,
the obtained $(4 \cdot 2^n) \times (4 \cdot 2^n)$ matrix is
distinguished by a single parameter,
where distinguished matrices are not similar matrices.
Then, we give a reasonable condition to fix the parameter.

In Sec.~II, we briefly review the basic properties of Cayley-Dickson
algebras, followed by some Lemmas concerning the representation matrices for
the right- and left-multiplication operators of the Cayley-Dickson algebras.
In Sec.~III, we first see how the Born-Infeld Lagrangian can be rewritten
using the determinant of the
$(4 \cdot 2^n) \times (4 \cdot 2^n)$ matrix, and then observe how the
obtained matrices are equivalently represented by imposing a reasonable
condition. In Sec.~IV, we give summary.

\section{Cayley-Dickson algebra}

\subsection{Basic properties}

In this subsection, we review the basic properties relevant to the present
study.

The Cayley-Dickson algebra $\alg_n$ over real numbers $\real$ is
the algebra structure given inductively in the following
way~\cite{schafer,moreno}.
Let
\begin{eqnarray*}
x = (x_1, x_2), \; y = (y_1, y_2) \;\; \mbox{in } \real^{2^n} =
\real^{2^{n-1}}
\times \real^{2^{n-1}},
\label{x}
\end{eqnarray*}
then addition and scalar multiplication are done
component-wise, and multiplication
is given by
\begin{equation}
xy = (x_1 y_1 - \bar{y}_2 x_2 , y_2 x_1  +  x_2 \bar{y}_1),
\label{xy}
\end{equation}
where $\bar{x} = (\bar{x}_1, - x_2)$.
Thus, for all $x,y,z$ in $\alg_n$, $\overline{xy} =
\bar{y} \bar{x}$, and
\begin{equation}
-(x,y,z) = (\bar{x}, y,z) = (x,\bar{y},z) = (x,y,\bar{z}),
\label{associator}
\end{equation}
where $(x,y,z) = (xy)z - x (yz)$, representing so called the associator.
The Euclidean norm and the inner product in $\real^{2^n}$ are given by
\begin{gather*}
\| x \|^2 = x \bar{x} = \bar{x} x = \| x_1 \|^2 + \| x_2 \|^2,\\
\langle x, y \rangle = \frac{1}{2} ( x \bar{y} + y \bar{x} ),
\end{gather*}
%
%
%
respectively.

Real numbers ($\real$), complex numbers (${\mathbb C}$),
quaternions ($\ham$), octonions (${\mathbb O}$), and sedenions (${\mathbb S}$)
correspond to
\begin{eqnarray*}
\real = \alg_0, \; {\mathbb C} = \alg_1, \; \ham = \alg_2, \;
{\mathbb O} = \alg_3, \; {\mathbb S} = \alg_4.
\end{eqnarray*}
For $n \leq 1$, $\alg_n$ is commutative: $xy = yx$; for $n \leq 2$, it
is associative: $(xy) z = x (yz)$; for $n \leq 3$, it is alternative: $(xx)y =
x(xy), x(yy) = (xy) y$; for all $n$, it is flexible: $x(yx) = (xy)x$.
Due to the flexibility, together with Eq.~(\ref{associator}),
we obtain 
\begin{eqnarray}
\langle x, yz \rangle = \langle x \bar{z}, y \rangle = \langle \bar{y} x, z
\rangle, \;\; \forall x,y,z \in \alg_n.
\label{inner}
\end{eqnarray}
Furthermore, we get
\begin{eqnarray}
\bar{x} (xy) = (y \bar{x}) x, \;\; \forall x,y \in \alg_n.
\label{flex}
\end{eqnarray}
Compared to the flexibility, Eq.~(\ref{flex}) is
less known. However, it is useful in obtaining the Lemma~4.

\subsection{Right and left multiplications}

To analyze the algebra structure of the Cayley-Dickson algebra, it is
convenient to introduce the operators $R_x$ and $L_x$ ($x \in
\alg_n$) representing the right and left multiplications as
$R_x, L_x: \alg_n \rightarrow \alg_n$ ($x \in \alg_n$) by
\begin{eqnarray*}
R_x (y) = yx, \; L_x (y) = xy, \; \forall y \in \alg_n.
\label{R}
\end{eqnarray*}
Apparently, $R_x$ and $L_x$ are linear, so that the corresponding
representation matrices
$M (x)$ and $\bar{M} (x)$ are given by
\begin{equation}
\left[
M (x) \right]_{ij} = \langle e_j, R_x (e_i) \rangle, \;
\left[
\bar{M} (x) \right]_{ij} = \langle e_j, L_x (e_i) \rangle,
\label{[Mx]}
\end{equation}
where $e_i$ ($i=0,1,\ldots,2^n-1$) represents
the canonical basis in
$\alg_n$, that is, $x = \sum_{i=0}^{2^n - 1} x^i e_i$, with $x^i \; \in
\real$ the component of $x$.
To express Eq.~(\ref{[Mx]}) in an index-free form, let us
introduce the bijective linear mapping $f: \alg_n \rightarrow \real^{2^n}$ by
\begin{align}
f (x) &= \left( \langle e_0, x \rangle, \ldots, \langle e_{2^n - 1}, x \rangle
\right) \nonumber \\
&= (x^0, \ldots, x^{2^n-1}).
\label{f}
\end{align}
Multiplying both sides of Eq.~(\ref{[Mx]}) by $y^i$, and by summing over $i$
from $0$ to $2^n - 1$, we obtain the $j$th component of the following
index-free identity:
\begin{align}
\forall x, y \in \alg_n, \;\;
f(yx) =
\begin{cases}
f(y) M (x), \\
f(x) \bar{M} (y).
\end{cases}
\label{fyx}
\end{align}
%
%
%

For later convenience,
we summarize the following properties concerning $f$:
\begin{subequations}
\begin{itemize}
\item
For $x \in \alg_n$,
\begin{eqnarray}
f(x) = 0 \Leftrightarrow x=0.
\label{fx}
\end{eqnarray}
\item
For ${\mathcal M}$ being $2^n \times 2^n$ matrix whose components are all
independent of $y$,
\begin{eqnarray}
\forall y \in \alg_n, \quad f(y) {\mathcal M} = 0
\Leftrightarrow {\mathcal M} = 0.
\label{forally}
\end{eqnarray}

\item
For all $x=(x_1,x_2)$ in $\alg_n = \alg_{n-1} \times \alg_{n-1}$,
\begin{eqnarray}
f (x) = (f(x_1), f(x_2)).
\label{fx=}
\end{eqnarray}
%
\end{itemize}
\end{subequations}
All of them are almost self-evident.

\subsection{Representation matrices}

In this subsection, we obtain some basic properties of $M(x)$ and $\bar{M}
(x)$.

\begin{description}
\item[Notation:] $B(x,y) = [ \bar{M} (y), M (x)]$, with $[X,Y]=XY-YX$.
\end{description}

To begin with, we deal with the case for $\alg_n \; (n \leq 2)$, where
the associativity is satisfied.
\begin{description}
\item[Lemma 1.]
For all $x, y$ in $\alg_n \; (n \leq 2)$

\begin{enumerate}
\renewcommand{\labelenumi}{\roman{enumi})}
\item
$M (xy) = M (x) M(y)$.

\item
$\bar{M} (\overline{xy}) = \bar{M} (\bar{x}) \bar{M} (\bar{y})$.

\item
$B (x,y) = 0$.

\end{enumerate}

\item[Proof.]
The associativity is satisfied for $\alg_n \; (n \leq 2)$.
Thus for all $x,y,z$ in $\alg_n \; (n \leq 2)$, we obtain
using Eq.~(\ref{fx}), the linearity of $f$, and Eq.~(\ref{fyx})
\begin{align*}
0
&= f( x(yz) - (xy)z ) \\
&= f(x(yz)) - f((xy)z) \\
&= f(x) M (yz) - f(xy) M (z) \\
&= f(x) [ M (yz) - M (y) M (z)],
\end{align*}
so that $M (yz) = M (y) M (z)$ by Eq.~(\ref{forally}). Thus, we
prove i).
By manipulating the right-hand side of the first line so as to be written as
in the forms $f(z)[\ldots]$ and $f(y) [ \ldots ]$, we obtain
$\bar{M} (xy) = \bar{M} (y) \bar{M} (x)$ and $[M (z), \bar{M} (x)] = 0$,
respectively. Thus, ii) and iii) are proven.
\hspace*{\fill} $\Box$

\end{description}
Since the maps $M (x), \bar{M} (\bar{x}): \alg_n \rightarrow
\real^{2^n}$ are bijective,
Lemmas~1.i) and ii) indicate the isomorphism
\begin{equation}
x \cong M (x) \cong \bar{M} (\bar{x}), \; \forall x \in \alg_n \; (n \leq 2),
\label{xcong}
\end{equation}
namely,
$M (x)$ and $\bar{M} (\bar{x})$, which are originally introduced as
representation matrices for the right and left multiplication
operators, respectively, also turn out to be the faithful representation
matrices for
$\alg_n \; (n \leq 2)$ itself. This is due to the associativity.
%
%

For $n \geq 3$, $\alg_n$ is not associative, so that the homomorphisms
corresponding Lemmas~1.\,i) and ii) are not satisfied, that is, a
faithful representation matrix is not realized for $\alg_n \; (n \geq 3)$.
Nonetheless, the matrices $M$ and $\bar{M}$ given by 
Eq.~(\ref{fyx}) are useful in analyzing the algebra structure even for
$\alg_n \; (n \geq 3)$.
\begin{description}
\item[Lemma~2.]
For all $x = (x_1, x_2)$ in $\alg_n = \alg_{n-1} \times \alg_{n-1}$, $M(x)$ and
$\bar{M} (x)$ can be decomposed into four blocks as
\begin{enumerate}
\renewcommand{\labelenumi}{\roman{enumi})}
%
\item
$M (x) =
\mat{ {M}(x_1) & \bar{M} (x_2)\\
- \bar{M} (\bar{x}_2) & {M} (\bar{x}_1) }$.

\item
$\bar{M} (x) =
\mat{ \bar{M} (x_1) & \eta \bar{M} (x_2) \\
- \eta {M} (x_2) & {M} (x_1) }$,

\end{enumerate}
where
$\eta = {\rm diag} \, (1,-1,-1,\ldots,-1)$.
%

\item[Proof.]
For all $x = (x_1, x_2)$ in $\alg_n = \alg_{n-1} \times \alg_{n-1}$,
we obtain from Eqs.~(\ref{fyx}), (\ref{xy}), 
(\ref{fx=}), and the linearity of $f$
\begin{align*}
&{} f(x) M (y)
\\
&= f(xy) \\
&= ( f(x_1 y_1 - \bar{y}_2 x_2), f(y_2 x_1 + x_2 \bar{y}_1)) 
\\
&=
\left(
f( x_1 y_1) - f( \bar{y}_2 x_2), f( y_2 x_1) + f( x_2 \bar{y}_1)
\right) \\
&=
\left(
f(x_1) {M} (y_1) - f(x_2) \bar{M} (\bar{y}_2),
f(x_1) \bar{M} (y_2) + f(x_2) {M} (\bar{y}_1)
\right)  \\
&=
\left(
f(x_1), f(x_2)
\right)
\mat{{M}(y_1) & \bar{M} (y_2)\\ - \bar{M} (\bar{y}_2) & {M} (\bar{y}_1)} \\
&= f(x)
\mat{{M}(y_1) & \bar{M} (y_2)\\ - \bar{M} (\bar{y}_2) & {M} (\bar{y}_1)},
\end{align*}
so we prove i) by Eq.~(\ref{forally}). In a similar way, we obtain ii) by
using $f(\bar{x}) = f(x) \eta$.
%
\hspace*{\fill} $\Box$

\end{description}

\begin{description}
\item[Lemma~3.]
For all $x,y$ in $\alg_n$,
\begin{enumerate}
\renewcommand{\labelenumi}{\roman{enumi})}
\item
${}^t \! M (x) = M (\bar{x})$.

\item
${}^t \! \bar{M} (x) = \bar{M} (\bar{x})$.

\item
$^t \! (\eta M (x)) = \eta \bar{M} (x)$.

\item
$B(x,y) = B(\bar{x},\bar{y}) = - {}^t \! B (x,y)$.

\end{enumerate}

\item[Proof.]
From Eqs.~(\ref{inner}) and (\ref{[Mx]}), we get
$[M(x)]_{ij} = \langle e_j, e_i x \rangle = \langle e_i x, e_j \rangle =
\langle e_i, e_j \bar{x} \rangle = [M(\bar{x})]_{ji}$, where use has been made
of $\langle x, y \rangle = \langle y, x \rangle$ for all $x,y$ in $\alg_n$, so
we prove i). In a similar way, we obtain ii).
To show iii), observe that $\bar{e}_i = \sum_{k=0}^{2^n-1} \eta_{ik} e_k$.
Thus, we have
$[ \eta \bar{M} (x) ]_{ij} = \sum_k \eta_{ik} \langle e_j, x e_k \rangle =
\langle e_j, x \bar{e}_i \rangle = \langle e_j e_i, x \rangle = \langle e_i,
\bar{e}_j x \rangle = \sum_k \eta_{jk} \langle e_i, e_k x \rangle = [ \eta M
(x) ]_{ji}$, so we obtain iii).
To show iv), first notice that $(x,y,z) = (\bar{x}, y, \bar{z})$ from
Eq.~(\ref{associator}). Then, for all
$x,y,z$ in $\alg_n$, we get
\begin{align*}
0 &= f ( (x,y,z) - (\bar{x}, y, \bar{z}) ) \\
&= f((xy)z) - f(x(yz)) + f(\bar{x} (y\bar{z})) - f((\bar{x}y) \bar{z}) \\
&= f(y) \left( [\bar{M} (x), M (z)] + [ M (\bar{z}), \bar{M} (\bar{x})]
\right),
\end{align*}
by Eq.~(\ref{fx}), the linearity of $f$, and Eq.~(\ref{fyx}). Using
Eq.~(\ref{forally}), we obtain
$B (z,x) - B (\bar{z}, \bar{x}) = 0$. On the other hand,
$B(\bar{z}, \bar{x})$ implies $- {}^t \! B (z,x)$ by i) and ii), so we
prove $B (z,x) = - {}^t  \! B (z,x)$.
\hspace*{\fill} $\Box$

\end{description}

In calculating the determinants of $M(x)$ and $\bar{M} (x)$ in the next
subsection, it is convenient to obtain some identities concerning the product
of
$M(x)$ and
$\bar{M} (x)$.

\begin{description}
\item[Notation:] For all $x$ in $\alg_n$,
$N(x) = M(x) M(\bar{x})$ and $\bar{N} (x) = \bar{M} (x) \bar{M} (\bar{x})$.
\end{description}

\begin{description}
\item[Corollary~1.]
For all $x$ in $\alg_n$,
\begin{eqnarray*}
{}^t \! N (x) = N (x), \;\; {}^t \! \bar{N} (x) = \bar{N} (x).
\end{eqnarray*}

\item[Proof.]
Follows from Lemmas~3.\,i) and ii).
\hspace*{\fill} $\Box$

\end{description}

\begin{description}
\item[Lemma~4.]
For all $x$ in $\alg_n$,
\begin{eqnarray*}
N(x) = N(\bar{x}) = \bar{N} (x) = \bar{N} (\bar{x}).
\end{eqnarray*}

\item[Proof.]
First of all, notice that for all $x,y$ in
$\alg_n$, $(yx)\bar{x} = (y\bar{x})x$, which is derived from $(y,x,\bar{x}) =
(y,\bar{x}, x)$ using Eq.~(\ref{associator}). Therefore, we get for all
$x,y$ in $\alg_n$,
\begin{align*}
0 &= f((yx)\bar{x}) - f((y\bar{x})x) \\
&= f(yx) M (\bar{x}) - f(y\bar{x}) M (x) \\
&= f(y) [ M(x) M(\bar{x}) - M (\bar{x}) M (x)],
\end{align*}
by Eq.~(\ref{fx}), the linearity of $f$, and Eq.~(\ref{fyx}).
Using Eq.~(\ref{forally}),
we
obtain
$N(x) = N (\bar{x})$ by Eq.~(\ref{fx}). Similarly, we obtain
$\bar{N} (x) =
\bar{N} (\bar{x})$ from $x (\bar{x} y) = \bar{x} (xy)$.
To show $N (\bar{x}) = \bar{N} (x)$, we use Eq.~(\ref{flex}), from which
\begin{align*}
0 &= f((y\bar{x}) x) - f(\bar{x} (xy)) \\
&= f(y\bar{x}) M (x) - f(xy) \bar{M} (\bar{x}) \\
&= f(y) [M (\bar{x}) M (x) - \bar{M} (x) \bar{M} (\bar{x}) ]
\end{align*}
for all $x,y$ in $\alg_n$, so we obtain $N (\bar{x}) = \bar{N} (x)$.
\hspace*{\fill} $\Box$

\end{description}

\subsection{Determinant}

We calculate the determinants of $M$ and $\bar{M}$.
\begin{description}
\item[Corollary~2.] For all $x$ in $\alg_n$,
\begin{eqnarray*}
| M (x) | = | M (\bar{x}) | = | \bar{M} (x) | = | \bar{M} (\bar{x}) |.
\label{Mx}
\end{eqnarray*}
\item[Proof.]
From Lemmas~3.\,i)-iii) and the determinant identities $| {}^t \! A | = | A
|$ and
$| A B | = |A| |B|$ for square matrices $A$ and $B$, we obtain Corollary~2.
\hspace*{\fill} $\Box$
\end{description}

Instead of calculating $| M (x) |$ directly, we evaluate
$| N(x) |$, where $| M (x) |^2 = | N (x) |$ is satisfied.
Using Lemmas~2, 3.\,iv), and 4, we can decompose $N (x)$ as
\begin{align}
N (x) &= \left(
\begin{matrix}
N (x_1) + \bar{N} (x_2) & B (x_1,x_2) \\
- B (\bar{x}_1, \bar{x}_2) & N (\bar{x}_1) + \bar{N} (\bar{x}_2)
\end{matrix}
\right) \nonumber \\
&=
\left(
\begin{matrix}
N (x_1) + N (x_2) & B (x_1,x_2) \\
- B (x_1, x_2) & N (x_1) + N (x_2)
\end{matrix}
\right)
\label{MM}
\end{align}
for $x= (x_1, x_2)$ in $\alg_n = \alg_{n-1} \times \alg_{n-1}$.

\begin{description}
\item[Lemma~5.]
For all $x = (x_1,x_2)$ in $\alg_n = \alg_{n-1} \times \alg_{n-1}$,
\begin{eqnarray*}
| M (x) | = | A(x_1,x_2) + i B(x_1,x_2) |,
\end{eqnarray*}
where $A(x_1, x_2) = N (x_1) + N (x_2)$.

\item[Proof.]
For all $x=(x_1,x_2)$ in $\alg_n = \alg_{n-1} \times \alg_{n-1}$,
${}^t \! A(x_1,x_2) = A (x_1,x_2)$ by Corollary~1, and
${}^t \! B (x_1,x_2) = - B (x_1,x_2)$ by Lemma~3.\,iv).
Thus, from
Eqs.~(\ref{MM}) and (\ref{b2}), we obtain
$| N (x) | = | A (x_1, x_2) + i B (x_1,x_2) |^2$, so that
$| M (x) | = \pm | A(x_1,x_2) + i B(x_1,x_2) |$.
Here, we show that the case
of the minus sign is not appropriate.
%
Putting $f(x) = (x^0,\ldots,x^{2^n - 1})$, we obtain
$M (x) = x^0 I_{2^n} + (x^0\text{-indep. term})$, where $I_{2^n}$ is the
$2^n$-dimensional unit matrix.
In this case, we find that
$A(x_1,x_2) = (x^0)^2 I_{2^{n-1}} + O (x^0)$, and that $B(x_1,x_2)$ is
independent of $x^0$.
Therefore, the coefficients of $(x^0)^{2^n}$ in $| M (x) | $ and
$| A (x_1,x_2) + i B (x_1,x_2) | $ are both unity,
indicating the inappropriateness of the case for the
minus sign.
\hspace*{\fill} $\Box$

\end{description}

\begin{description}
\item[Lemma~6.]
For all $x$ in $\alg_n \; (n =1,2, 3)$,
\begin{eqnarray*}
| M (x) | = \| x \|^{2^n}.
\end{eqnarray*}

\item[Proof.]
For all $x_1,x_2$ in $\alg_{n-1}$ ($n \leq 3$),
we find that
$B(x_1,x_2) = 0$ and
$N (x_1) = \| x_1 \|^2 I_{2^{n-1}}$ by Lemma~1, so that the latter leads to
$A(x_1,x_2) = ( \| x_1 \|^2 + \| x_2 \|^2 ) I_{2^{n-1}}$.
Thus, for $x=(x_1,x_2)$ in $\alg_n = \alg_{n-1} \times
\alg_{n-1}$ ($n \leq 3$),
$| M (x) | = | A(x_1,x_2) | = | \| x \|^2 I_{2^{n-1}} | = \| x \|^{2^n}$
by Lemma~5.
\hspace*{\fill} $\Box$

\end{description}
For $x_1, x_2 \in \alg_{n-1} \; (n \geq 4)$, $B(x_1,x_2)$ is not in general
vanishing due to the non-associativity, which implies that it may  be rather
complicated to calculate $| M (x) |$ for
$x
\in \alg_n
\; (n \geq 4)$.
As the simplest one of them, where $x \in \alg_4$, $| M (x) |$ is calculated
to be shown in Appendix~A.
Fortunately, in the next section, we will only deal with the case where the
associativity is satisfied, that is,
$B(x_1,x_2) = 0$. In this case, $| M (x) |$ turns out to be the same form as
in Lemma~6.

\begin{description}
\item[Notation:]
$\alg_n^{(a)} = \{ (x_1,x_2) \in \alg_{n-1}^{(a)} \times
\alg_{n-1}^{(a)} | B (x_1,x_2) = 0 \}$, with $\alg_0^{(a)} = \alg_0$.
\end{description}

\begin{description}
\item[Corollary~3.]
For all $x$ in $\alg_n^{(a)} \; (n \geq 1)$,
\begin{eqnarray*}
| M (x) | = \| x \|^{2^n}.
\end{eqnarray*}
\item[Proof.] For all $x$ in $\alg_{n-1}^{(a)}$, we find from Eq.~(\ref{MM})
that
$N(x) = \| x \|^{2} I_{2^{n-1}}$ by induction.
Thus for
$x= (x_1,x_2)$ in
$\alg_n^{(a)} = \alg_{n-1}^{(a)} \times \alg_{n-1}^{(a)}$,
$A(x_1,x_2) = ( \| x_1 \|^2 + \| x_2 \|^2) I_{2^{n-1}} = \| x \|^2 
I_{2^{n-1}}$.
Therefore, we obtain for $x \in \alg_n^{(a)}$,
$| M (x) | = | \| x \|^2 I_{2^{n-1}} | = \| x \|^{2^n}$ by Lemma~5.
\hspace*{\fill} $\Box$

\item[Remark~1:]
$\alg_n^{(a)} = \alg_n \;\; (n \leq 3)$.

\end{description}

\section{Born-Infeld Lagrangian}

\subsection{8-dimensional phase space}

In this subsection, we rewrite the Born-Infeld Lagrangian using the $8
\times 8$ matrix.
Originally, the Born-Infeld Lagrangian ${\mathcal L}$ is given using the $4
\times 4$ matrix by~\cite{Born}
\begin{eqnarray*}
{\mathcal L} = (-|H_4|)^{1/2}, \;\; H_4 = g + bF,
\label{L}
\end{eqnarray*}
where $g$ and $F$ 
represent the
Minkowski metric and the electromagnetic field strength tensor, respectively;
and $b$ is a parameter characteristic of the Born-Infeld dynamics.
In a frame where the magnetic field ${\bf B}$ is negligible,
the Born-Infeld Lagrangian implies that
the electric field ${\bf E}$ should be bounded above as $| {\bf E} | < | b
|^{-1}$,
because ${\mathcal L} \rightarrow (1-b^2 {\bf E}^2)^{1/2}$ for
${\bf B} \rightarrow 0$.

From the determinant identity for square matrices $A, B$ (see Appendix~B)
\begin{equation}
\left|
\begin{array}{cc}
A & -B\\
B & -A
\end{array}
\right| = | A + B | \, | {}^t \! B - {}^t \! A |,
\label{A-B}
\end{equation}
${\mathcal L}$ can be rewritten as
\begin{equation}
{\mathcal L} = | H_8 (0) |^{1/4}, \;\;
H_8 (0) = \mat{bF & - g \\ g & -bF},
\label{L=}
\end{equation}
where use has been made of ${}^t \!  g = g$ and ${}^t \! F = - F$
(notice that the determinant identity $| H_4 |^2 = | H_8 (0) |$ holds for any
symmetric
$g$ and any antisymmetric $F$).
The obtained $8 \times 8$ matrix is antisymmetric, so that its elements can
be written as the commutator of 8-dimensional variables of the form
\begin{eqnarray*}
\left[
H_8 (0)
\right]_{nm} = [X_n, \, X_m] \;\;\; (n,m=1,\ldots,8).
\label{H8}
\end{eqnarray*}
If one puts $X_n$ as~\cite{schuller}
\begin{eqnarray*}
(X_1, \ldots, X_8) = \sqrt{i \beta }
\, ( x_\mu /
\beta ; \,  p_\mu ),
\label{X1}
\end{eqnarray*}
where $\beta = b / e$, with $e$ the elementary
charge, then $x_\mu$ and $p_\mu$ should satisfy the following commutation
relations:
\begin{eqnarray*}
[x_\mu,  x_\nu] = -i e \beta^2 F_{\mu \nu},
\; [x_\mu,  p_\nu] = i g_{\mu \nu}, \;
[p_\mu,  p_\nu] = i e F_{\mu \nu}.
\label{xmu}
\end{eqnarray*}
From the last commutation relation, $p_\mu$ can be identified with the
canonical momentum, so that from the second one, $x_\mu$ can be regarded as
the position. Thus from the first one, non-commutative spacetime is
suggested.
Eventually, the elements of $H_8 (0)$ can be given by the 8-dimensional phase
space variables with non-commutative spacetime.
As $b$ tends to be vanishing, which corresponds to the classical limit, the
non-commutativity of the spacetime is suppressed, recovering the commutative
spacetime.

Here, we have a problem whether the matrix $H_8$ such that $| H_8 | = |H_4|^2$
is uniquely determined or not (except those obtained through a similarity
transformation).
Even if $H_8$ is restricted to such that the four blocks of $4 \times 4$ matrix
are all given by the linear combination of $g$ and $F$, it will be found
that the answer is still negative.

Consider the matrix $H_8 (\theta)$ such that
\begin{equation}
H_8 (\theta) = \mat{bF + \sin \theta \cdot g & - \cos \theta \cdot g \\
\cos \theta \cdot g & - bF + \sin \theta \cdot g},
\label{H8theta}
\end{equation}
whose four blocks of $4 \times 4$ matrix are indeed given by
the linear combination of $g$ and $F$
[notice that
$H_8 (0)$ in
Eq.~(\ref{L=}) is given by $H_8 (\theta = 0)$].
In this case, we find from the second identity in Eq.~(\ref{c7}), that
$| H_8 (\theta) |$ is independent of $\theta$, namely
\begin{align}
{}&| H_8 (\theta) | \nonumber \\
&=
\left| \cos^2 \theta \cdot g^2 + g (-bF + \sin \theta \cdot g)
g^{-1} (bF + \sin \theta \cdot g) \right| \nonumber \\
&=
\left| \cos^2 \theta \cdot g^2 - b^2 g F g^{-1} F + \sin^2 \theta
\cdot g^2 \right| \nonumber \\
&= | H_8 (0) |.
\label{|H8|}
\end{align}
Furthermore, we find that
$H_8 (d \theta)$ and $H_8
(0)$ are not similar matrices as follows.
Suppose that
$H_8 (d \theta)$
can be written as
$S H_8 (0) S^{-1}$.
Assume that $S$ is analytic around $\theta = 0$, where $S$ can be expanded as
$S = S_0 +S_1 d \theta + O (d \theta^2 )$.
Comparing the terms proportional to $d \theta^0$ in
$H_8 ( d \theta ) = S H_8 (0) S^{-1}$, we obtain $H_8 (0) = S_0 H_8 (0)
S_0^{-1}$.
From the terms proportional to $d \theta$, it is required that
\begin{equation}
\left. \partial H_8 (\theta ) / \partial \theta \right|_{\theta = 0}
= [T, H_8 (0)].
\label{partialH}
\end{equation}
where $T = S_1 S_0^{-1}$.
While the trace in the left-hand side of Eq.~(\ref{partialH})
is
$2 \, {\rm Tr}
\, g \; (\neq 0)$, the trace in the right-hand side is vanishing, due to the
cyclic property of the trace.
Thus, it is found that
$H_8 (d \theta)$ cannot be obtained from the similarity transformation of
$H_8 (0)$.

In obtaining $H_8 (\theta)$ in Eq.~(\ref{H8theta}), the
Cayley-Dickson algebras (in this case, quaternions and octonions) play a useful
role.
To see this, notice first that $F$ can be written using the
representation matrices for quaternions
$q, q' \in\ham$ as (see Appendix~C)
\begin{subequations}
\label{F}
\begin{align}
{F}_{\mu \nu} &= \left[ M (q) \right]_{\mu \nu} + \frac{1}{2}
\left[ {\rm Ad} (q') \right]_{\mu \nu}, \\
{F}^{\mu \nu} &= \left[ \bar{M} (\bar{q}) \right]_{\mu \nu} + \frac{1}{2}
\left[ {\rm Ad} (q') \right]_{\mu \nu},
\end{align}
\end{subequations}
where ${\rm Ad} (x) \equiv \bar{M} (x) - M (x)$, and the components of $q$
and $q'$ are
\begin{eqnarray*}
f (q) = (0,{\bf E}), \;\; f (q')  = (0, {\bf E} - {\bf B}).
\end{eqnarray*}
Considering that $M$ is linear and that
$M (e_0)$ corresponds to the unit matrix, we obtain from Eq.~(\ref{F})
\begin{subequations}
\label{action}
\begin{align}
\delta_{\mu \nu} + b
{F}_{\mu \nu} &= \left[ M (e_0 + bq) \right]_{\mu \nu} + \frac{b}{2}
\left[ {\rm Ad} (q') \right]_{\mu \nu}, \\
\delta^{\mu \nu} + b
{F}^{\mu \nu} &= \left[ \bar{M} (\overline{e_0 + bq}) \right]_{\mu \nu} +
\frac{b}{2}
\left[ {\rm Ad} (q') \right]_{\mu \nu},
\end{align}
\end{subequations}
where $\delta_{\mu \nu}$ represents the 4-dimensional Euclidean metric, and
$b$ is an arbitrary real number.

For the time being, we consider the limiting case of $q' \rightarrow 0$, for
simplicity. Then we examine whether the resultant determinant identity may be
applied to any antisymmetric $F$ or not. This is a practical way, if
successful, to achieve a desired result. 
For $q' \rightarrow 0$, where ${\rm Ad} (q') \rightarrow 0$, we obtain from
Eq.~(\ref{F}), Lemma~6, and Corollary~2, the determinant identity
\begin{eqnarray*}
| \delta + b \hat{F} | = \| h \|^4,
\label{h}
\end{eqnarray*}
%
where $\hat{F} = \left. F \right|_{{\bf B} \rightarrow {\bf E}}$, and
\begin{equation}
h = e_0 + b q \in \ham.
\label{h=}
\end{equation}
Notice that the Euclideanized Born-Infeld Lagrangian can be obtained by
replacing
${\bf E}$ in the original Lagrangian by $i {\bf E}$ (except the overall minus
sign), that is
\begin{eqnarray*}
| \delta + b F | =
- | g + b \check{F} |,
\end{eqnarray*}
where $\check{F} = \left. F \right|_{{\bf E} \rightarrow i{\bf E}}$,
so that we can deal with the Euclideanized Lagrangian instead of the
original Lagrangian.

In octonionizing $h$ in Eq.~(\ref{h=}) as
$h \rightarrow o = (h_1,h_2)$ $ \in {\mathbb O} = \ham \times \ham$, such
that
\begin{equation}
\| h \|^2 = \| o \|^2 \; (= \| h_1 \|^2 + \| h_2 \|^2),
\label{|h|}
\end{equation}
the matrix form of $M (x)$ in Lemma~2 is useful, namely
\begin{equation}
M (o) = \mat{M (h_1) & \bar{M} (h_2) \\
- \bar{M} (\bar{h}_2) & M (\bar{h}_1)}.
\label{Mo}
\end{equation}
Under the condition that the four blocks of $4 \times 4$ matrix in the
right-hand side of Eq.~(\ref{Mo}) are given by the linear combination of
$\delta$ and $\hat{F}$ (this means that the $4 \times 4$ matrices are rank-2
tensors of the same type as $\delta$ and $\hat{F}$), it is required that
either
\begin{equation}
h_2 = \bar{h}_2, \; \text{or } h_1 = \bar{h}_1;
\label{h1}
\end{equation}
otherwise, both contravariant component $\hat{F}_{\mu \nu}$ and
covariant component $\hat{F}^{\mu
\nu}$ would appear in the elements in $M (o)$ in Eq.~(\ref{Mo}).
Considering the condition of Eq.~(\ref{|h|}), we can set
either
\begin{subequations}
\label{h1=}
\begin{eqnarray}
f (h_1) = (\sin \theta, b {\bf E}), \;\;
f (h_2) = (- \cos \theta, {\bf 0}),
\end{eqnarray}
or
\begin{eqnarray}
f (h_1) = (\sin \theta, {\bf 0}), \;\;
f (h_2) = (- \cos \theta, - b {\bf E}).
\end{eqnarray}
\end{subequations}
Recalling that $| \delta + b \hat{F} |^2 = \| h \|^8 = \| o \|^8 = | M(o) |$,
and using Eq.~(\ref{h1=}),
we obtain
\begin{equation}
| \delta + b \hat{F} |^2 = | \hat{H}_8 (\theta) | = | \hat{H}'_8 (\theta) |,
\label{delta}
\end{equation}
where
\begin{align*}
\hat{H}_8 (\theta) &=
\mat{b \hat{F} + \sin \theta \cdot \delta & - \cos \theta \cdot \delta \\
\cos \theta \cdot \delta & - b \hat{F} + \sin \theta \cdot \delta }, \nonumber
\\
\hat{H}'_8 (\theta) &=
\mat{\sin \theta \cdot \delta  &  b \hat{F} - \cos \theta \cdot \delta \\
b \hat{F} + \cos \theta \cdot \delta & \sin \theta \cdot \delta}.
\label{hatH}
\end{align*}
Notice that $\hat{H}'_8 (\theta)$ can be derived from $\hat{H}_8 (\theta)$
through a similarity transformation (more strictly, orthogonal
transformation), namely,
$\hat{H}'_8 (\theta) = S
\hat{H}_8 (\theta) S^{-1}$, where
$S= \frac{1}{\sqrt{2}} \left( \begin{smallmatrix}
1 & -1 \\ 1 & 1 \end{smallmatrix} \right)$.
Thus, we will concentrate on the case of $\hat{H}_8 (\theta)$, simply ignoring
the case of $\hat{H}'_8 (\theta)$.

Considering that the identity of $|H_4|^2 = |H_8 (0)|$ holds
for any symmetric $g$ and any antisymmetric $F$,
we expect that the identity of Eq.~(\ref{delta}) should hold for any symmetric
$\delta$ and any antisymmetric $\hat{F}$. Actually, using
Eq.~(\ref{c7}), we find that this is found to be true, as was shown in
Eq.~(\ref{|H8|}). Eventually, Eq.~(\ref{H8theta}) is obtained.

One may wonder why
the identity of Eq.~(\ref{delta}) holds for
any symmetric $\delta$ and any antisymmetric $\hat{F}$, and suspect that
the identity is realized only by chance. However, the success in
realizing the identity can be understood by interpreting $\hat{H}_8
(\theta)$ and
$\hat{H}'_8 (\theta)$ as being composed of the direct product between
the 2-dimensional representation matrix for a parameterized complex number
[in this case, the representation matrix $\left( \begin{smallmatrix} \sin
\theta & -
\cos
\theta
\\
\cos \theta & \sin \theta \end{smallmatrix} \right)$ for the complex
number ($\sin
\theta - i
\cos \theta$)]
and
the original 4-dimensional symmetric tensor $\delta$ (or the antisymmetric
tensor
$\hat{F}$, as the case may be),
rather than by simply regarding them as the original
$8$-dimensional representation matrices for the octonionic multiplication.
This interpretation will manifest itself from the following subsection on.

\subsection{16 dimensional space}

Before generalizing the determinant identity in the previous subsection, we
``sedenionize''
$h$ in Eq.~(\ref{h=}) as
$h
\rightarrow s = (o_1, o_2) = (h_1,h_2,h_3,h_4)$ $ \in \alg_4 = {\mathbb
O}^2 = \ham^4$, such that
\begin{eqnarray}
\| h \|^2 = \| s \|^2 \; (= {\textstyle \sum_{i=1}^4 \| h_i \|^2}).
\label{|h|2}
\end{eqnarray}
Using Lemma~2 twice, we can express $M (s)$ in terms of $h_1, \ldots, h_4$
as
{
\renewcommand{\arraystretch}{1.2}
\begin{equation}
M (s) = \left(
\begin{array}{cc|cc}
M (h_1) & \bar{M} (h_2) & \bar{M} (h_3) & \eta \bar{M} (h_4) \\
-\bar{M} (\bar{h}_2) & M(\bar{h}_1)  & -\eta M (h_4) & M (h_3) \\
\hline
-\bar{M} (\bar{h}_3) & \eta \bar{M} (h_4) & M (\bar{h}_1) & - \bar{M}
(h_2)
\\
-\eta M (h_4) & -M (\bar{h}_3) &  \bar{M} (\bar{h}_2) & M (h_1)
\label{Ms}
\end{array} \right).
\end{equation}
}

Assume that the 16 blocks of $4 \times 4$ matrix in the
right-hand side of Eq.~(\ref{Ms}) are given by the linear combination of
$\delta$ and $\hat{F}$ (rank-2 tensors of the same type as $\delta$ and
$\hat{F}$). Then, it is required that
\begin{equation}
(h_1 - \bar{h}_1) (h_2 - \bar{h}_2) = 0,
\label{h2=}
\end{equation}
namely, $h_1 = \bar{h}_1$ or $h_2 = \bar{h}_2$ as in the same form as
Eq.~(\ref{h1}), and that
\begin{equation}
h_3 - \bar{h}_3 = h_4 = 0.
\label{h3}
\end{equation}
From Eq.~(\ref{h3}), it is found that $s$
belongs to
\begin{eqnarray*}
s \in \alg_4^{(a)}.
\label{s}
\end{eqnarray*}
This is
because $\bar{M} (o_2)$, where $o_2 = (h_3, h_4) = (\bar{h}_3,0)$ in
${\mathbb O} =
\ham^2$, turns out to be proportional to a unit matrix,
so that
$B(o_1,o_2) = [\bar{M} (o_2), M (o_1)] = 0$. Thus in this case,
$| M (s) | = \| s \|^{16}$ by Corollary~3.
As in the previous case of octonionizing $h$,
the matrix obtained under the condition of $h_1 = \bar{h}_1$ can be
derived from the analogous orthogonal transformation of another matrix
obtained under the condition of
$h_2 =
\bar{h}_2$. In this
sense, there is no loss of generality that we can choose
$h_2 = \bar{h}_2$. Thus, we can set $h_i$ ($i=1,2,3$) as
%
\begin{align}
f (h_1) = (c_0, b {\bf E}), \;
f (h_2) = (c_1, {\bf 0}), \;
f (h_3) = (c_2, {\bf 0}),
\label{h1-4}
\end{align}
%
%
where
$\sum_{i=0}^2 c_i^2 = 1$ so as to guarantee Eq.~(\ref{|h|2}).
Recalling that $| \delta + b \hat{F} |^4 = \| h \|^{16} = \| s \|^{16} = | M
(s) |$. We obtain from Eqs.~(\ref{Ms}) and (\ref{h1-4}) the identity
\begin{equation}
| \delta + b \hat{F} |^4 = | \hat{H}_{16} (c_0,c_1,c_2) |,
\label{delta+b}
\end{equation}
where
\begin{eqnarray*}
\lefteqn{
\hat{H}_{16} (c_0,c_1,c_2)} \nonumber \\
& = & \left(
\begin{array}{cccc}
b \hat{F} + c_0 \delta & c_1 \delta & c_2 \delta & 0 \\
-c_1 \delta & - b \hat{F} + c_0 \delta & 0 & c_2 \delta \\
-c_2 \delta & 0 & - b \hat{F} + c_0 \delta & - c_1 \delta \\
0 & -c_2 \delta & c_1 \delta & b \hat{F} + c_0 \delta
\end{array}
\right) \nonumber \\
&=&
{\rm diag} \, (b\hat{F}, -b \hat{F}, - b \hat{F}, b \hat{F}) + M (c^{(2)})
\otimes
\delta,
\label{H16}
\end{eqnarray*}
with the components of $c^{(2)} \in \ham$ given by
\begin{eqnarray*}
f (c^{(2)} ) = (c_0,c_1,c_2,0).
\label{c}
\end{eqnarray*}

Now we examine whether or not $\hat{H}_{16} (c_0', c_1', c_2')$ and
$\hat{H}_{16} (c_0, c_1, c_2)$ are similar matrices.
Denote $\hat{H}_{16} (\sin \theta, \cos \theta \cos \phi, \cos \theta \sin
\phi)$ by $H_{16} ( \theta, \phi)$.
As in the case of $H_8 (\theta)$, it is found that $H_{16} (\theta + d
\theta,
\phi)$ and $H_{16} (\theta, \phi)$ are not similar matrices.
This is verified in the way analogous to that done for $H_8 (\theta)$.
Suppose that $H_{16} (\theta + d \theta, \phi)$ and $H_{16} (\theta, \phi)$
are similar matrices. Then, there should exist $T_\theta$ such that
$\partial H_{16} (\theta, \phi) / \partial \theta = [T_\theta, H_{16}
(\theta, \phi)]$.
The trace of the right-hand side is vanishing, while that of the
left-hand side is $4 \cos \theta \,{\rm Tr} \delta $.
Thus, for $\cos \theta \neq 0$, $H_{16} (\theta + d \theta,
\phi)$ and $H_{16} (\theta, \phi)$ are not similar matrices (for $\cos
\theta = 0$, it is rather difficult to show that they are not similar
matrices).
However, it should be noticed that $H_{16} (\theta, \phi')$ and $H_{16}
(\theta, \phi)$ are similar matrices.
Suppose that $H_{16} (\theta, \phi')$ and $H_{16}
(\theta, \phi)$ are similar matrices. Then, there should exist the generator
$T_\phi$ such that
$\partial H_{16} (\theta, \phi) / \partial \phi = [T_\phi, H_{16}
(\theta, \phi)]$.
By direct calculation, $T_\phi$ can be
chosen as
$T_\phi=\left( \begin{smallmatrix} 0 & J \\ J & 0
\end{smallmatrix} \right)$, where $J = \frac{1}{2} \left( \begin{smallmatrix}
0 & I_4 \\ -I_4 & 0 \end{smallmatrix} \right)$ ($I_4$ is the 4-dimensional
unit matrix, see also Proposition~2).
In this case, we can write
$H_{16} (\theta, \phi) = S H_{16} (\theta, 0) S^{-1}$, where $S=
\exp (\phi \cdot  T_\phi) = \cos (\phi/2) \cdot I_{16}  + \sin
(\phi / 2) \cdot \left( \begin{smallmatrix} 0 & 2J \\ 2J & 0
\end{smallmatrix} \right)$.

Notice that $T_\phi$ is chosen as antisymmetric, so that $S$ turns out
to be an orthogonal matrix.
Furthermore, it may be interesting to point out that
$(-2T_\phi)$ corresponds to $C \otimes I_4$, where $C$ is the charge
conjugation matrix ($C \gamma_\mu C^{-1} = - {}^t \gamma_\mu$) in the
Dirac and Majorana representations~\cite{zuber}.
While $H_{16} (\theta, \phi)$ has a $2 \pi$ periodicity with respect to
$\phi$,
$S$ has a $4 \pi$ periodicity.
At first,
one may wonder why the spinor representation (more strictly,
the underlying Clifford algebra) is realized in the sedenion, by considering
that the Clifford algebra is
associative, while the sedenion is not, in general, associative.
However, the sedenion which is composed of four quaternions satisfying both
Eqs.~(\ref{h2=}) and (\ref{h3}) forms an associative algebra.
Thus, the appearance of the spinor representation is of no wonder.

Considering that Eq.~(\ref{delta}) holds for any symmetric
$\delta$ and antisymmetric
$\hat{F}$, we expect that Eq.~(\ref{delta+b}) should also hold for any
symmetric
$\delta$ and antisymmetric $\hat{F}$. This is found to be true, which will be
proven in the next subsection.

\subsection{$2^n N$ dimensional space}

In this subsections, we generalize the previous
determinant identity.
\begin{description}
\item[Notation:] Pauli matrices $\sigma_1 = \left( \begin{smallmatrix} 0 & 1
\\ 1 & 0 \end{smallmatrix} \right)$,
$\sigma_2 = \left( \begin{smallmatrix} 0 & -i
\\ i & 0 \end{smallmatrix} \right)$, 
$\sigma_3 = \left( \begin{smallmatrix} 1 & 0
\\ 0 & -1 \end{smallmatrix} \right)$.
\end{description}

\begin{description}
\item[Proposition~1.]
For any $N$-dimensional square matrix
\begin{eqnarray*}
Z \; = Z_+ + Z_- \;\; [\mbox{with } Z_\pm = (Z \pm {}^t \! Z) / 2],
\end{eqnarray*}
there exists a
$2^n N$-dimensional matrix $\mathcal{Z} (c_0,\ldots,c_{n})$ parameterized by
$(c_0, \ldots, c_{n}) \in \real^{n+1}$ with
\begin{equation}
{\textstyle \sum_{i=0}^{n} c_i^2= 1},
\label{ci2}
\end{equation}
such that
\begin{eqnarray*}
| Z |^{2^n} = | \mathcal{Z} (c_0, \ldots, c_{n}) |,
\end{eqnarray*}
and that the $4^n$ blocks of $N \times N$ matrix in ${\mathcal Z} (c_0,
\ldots,c_{n})$ are given by the linear combination of $Z_+$ and
$Z_-$ (this means that they are rank-2 tensors of the same
type as $Z$).
Based on the representation matrix $M(x)$ for the Cayley-Dickson algebras,
${\mathcal Z} (c_0,
\ldots, c_{n})$ can be written as
\begin{equation}
{\mathcal Z} (c_0,\ldots,c_{n}) = Z_-^{(n)} + M (c^{(n)}) \otimes Z_{+},
\label{mathZ}
\end{equation}
where $Z_-^{(n)}$ and $c^{(n)} \in \alg_n$ are given inductively by
\begin{eqnarray*}
Z_-^{(n)} = \sigma_3 \otimes Z_-^{(n-1)},
\;\; \text{with } Z_-^{(0)} = Z_{-},
\label{Dn}
\end{eqnarray*}
and
\begin{eqnarray*}
c^{(n)} = ( c^{(n-1)}, c_{n} e_0) \in \alg_{n-1} \times \alg_{n-1},
\label{cn}
\end{eqnarray*}
with $c^{(0)} = c_0$, and $e_0 = (1,0,\ldots,0) \in \alg_{n-1}$, denoting the
unit element.
For example, $c^{(1)} = (c_0,c_1)$, $c^{(2)} = (c_0,c_1,c_2,0)$, $c^{(3)} =
(c_0,c_1,c_2,0,c_3,0,0,0)$.
\item[Remark~2:]
%
$c^{(n)} \in \alg_n^{(a)}$ for all $n$, so that $c^{(n)}$ forms an
associative algebra.

\item[Proof.] We show $| Z |^{2^n} = | {\mathcal Z} |$ in the following way.
Denote
$F_n$ and
$G_n$ by
\begin{align*}
F_n &= B_n D_n B_n^{-1} A_n, \\
G_n &= B_n A_n B_n^{-1} D_n,
\end{align*}
where
\begin{align*}
B_n &= I_{2^n} \otimes (c_{n+1} Z_+), \\
A_n &= Z_-^{(n)} + M (c^{(n)} ) \otimes Z_+ \; [ = {\mathcal Z} (c_0, \ldots,
c_n)],
\\ D_n &= - Z_-^{(n)} + {}^t \! M (c^{(n)} ) \otimes Z_+,
\end{align*}
with $I_{2^n}$ the $2^{n}$-dimensional unit matrix.
To begin with, we show the following identities by induction:
\begin{eqnarray}
F_n = G_n = I_{2^n} \otimes
J_n,
\label{Fn}
\end{eqnarray}
where
$J_n =
\sum_{i=0}^{n} (c_i Z_+)^2 - Z_+^2 (Z_+^{-1} Z_-)^2
$ [notice at the present stage, that we have not yet used the condition of
Eq.~(\ref{ci2})].
For $n=0$,
$B_0 = c_1 Z_+$, $A_0 = Z_- + c_0 Z_+$, $D_0=-Z_- + c_0 Z_+$, so that
$F_0 = G_0 = (c_0 Z_+)^2 - Z_+^2 (Z_+^{-1} Z_-)^2$, which guarantees
Eq.~(\ref{Fn}) for $n=0$.
Suppose that Eq.~(\ref{Fn}) holds for $n=k$, namely,
$F_k = G_k = I_{2^k} \otimes J_k$. Then, $F_{k+1}$ and
$G_{k+1}$ can be written using the identities
%
%
%
\begin{align*}
B_{k+1} &= \frac{c_{k+2}}{c_{k+1}} \mat{B_k & 0 \\ 0 & B_k}, \\
A_{k+1} &= \mat{A_k & B_k \\ - B_k & D_k}, \\
D_{k+1} &= \mat{D_k & - B_k \\ B_k & A_k},
\end{align*}
as
\begin{align*}
F_{k+1} &= \mat{F_k + B_k^2 & 0 \\ 0 & G_k + B_k^2}, \\
G_{k+1} &= \mat{G_k + B_k^2 & 0 \\ 0 & F_k + B_k^2}.
\end{align*}
Noticing that $B_k^2 = I_{2^k} \otimes (c_{k+1} Z_+)^2$, we get $F_k + B_k^2 =
G_k + B_k^2 = I_{2^k} \otimes J_{k+1}$, so that Eq.~(\ref{Fn}) holds for
$n = k+1$. Thus, Eq.~(\ref{Fn}) holds for any $n$.
The next thing is to calculate $| \mathcal Z |$.
Recalling that
${\mathcal Z} (c_0, \ldots, c_n) = A_{n} = \left( \begin{smallmatrix}
A_{n-1} & B_{n-1} \\ - B_{n-1} & D_{n-1} \end{smallmatrix} \right)$,
we can rewrite
$| {\mathcal Z} (c_0,\ldots, c_n) |$ using the second
identity in Eq.~(\ref{c7}) as
\begin{align*}
| {\mathcal Z} (c_0,\ldots,c_n) |
&=| B_{n-1}^2 + B_{n-1} D_{n-1} B_{n-1}^{-1} A_{n-1} |
\\
&=| I_{2^{n-1}} \otimes (c_n Z_+)^2 + F_{n-1} | \\
&=| I_{2^{n-1}} \otimes (c_n Z_+)^2 + I_{2^{n-1}} \otimes J_{n-1} | \\
&=| I_{2^{n-1}} \otimes J_n |.
\end{align*}
Now we impose the condition of Eq.~(\ref{ci2}).
Under the condition, $J_n$ turns out to be independent of
$c_i$, and can be factorized as
$J_n = Z_+^2 (I_N + Z_+^{-1} Z_-) (I_N - Z_+^{-1} Z_-)$, where $I_N$ is the
$N$-dimensional unit matrix.
Using the determinant identity for square matrices $A$ and $B$:
$| I_{n} \otimes (AB) | = | AB |^n = ( | A | | B |)^n$, we
find that
\begin{align}
| {\mathcal Z} (c_0,\ldots,c_n) |
&=
\left(
| Z_+ |^2 | I_N + Z_+^{-1} Z_- | | I_N - Z_+ Z_- |
\right)^{2^{n-1}} \nonumber \\
&=
\left( \left| Z_+ + Z_- \right|
\left| Z_+ - Z_- \right| \right)^{2^{n-1}} \nonumber \\
&=
\left( \left| Z_+ + Z_- \right|
\left| {}^t \!Z_+ - {}^t \! Z_- \right| \right)^{2^{n-1}}.
\label{Zc0}
\end{align}
Therefore, we obtain
\begin{eqnarray*}
| {\mathcal Z} (c_0, \ldots, c_n) | =
\left|
Z_+ + Z_- \right|^{2^n} = | Z |^{2^n},
\end{eqnarray*}
due to ${}^t \! Z_+ = Z_+$ and ${}^t \! Z_- = - Z_-$.
\hspace*{\fill} $\Box$

\item[Remark~3:] The identity of Eq.~(\ref{Zc0}) holds for arbitrary square
matrices
$Z_+$ and $Z_-$; neither the condition of ${}^t \! Z_+ = Z_+$ nor ${}^t \! Z_-
= - Z_-$ is necessary for Eq.~(\ref{Zc0}).

\end{description}

Now we examine whether or not ${\mathcal Z} (c_0, \ldots, c_n)$ and
${\mathcal Z} (c_0', \ldots, c_n')$ are similar matrices.

\begin{description}

\item[Notation:] ${\mathcal Z}_\rho (\theta; \phi_1, \ldots, \phi_{n-1})$
denotes ${\mathcal Z} (c_0, \dots, c_n)$ with $c_i$ ($i=0,\ldots, n$)
substituted by the following $(n+1)$-dimensional unit polar coordinates:
\begin{align*}
c_1 &= \cos \theta \, \cos \phi_{n-1} \,\cos \phi_{n-2} \ldots \cos \phi_2 \,
\cos \phi_1,\\
c_2 &= \cos \theta \, \cos \phi_{n-1} \,\cos \phi_{n-2} \ldots \cos \phi_2 \,
\sin \phi_1,\\
& \vdots \\
c_{n-1} &= \cos \theta \,\cos \phi_{n-1} \,\sin \phi_{n-2}, \\
c_n &= \cos \theta \, \sin \phi_{n-1}, \\
c_0 &= \sin \theta.
\end{align*}
Denote the $\phi_i$-dependence of ${\mathcal Z}_\rho (\theta; \phi_1,
\ldots,
\phi_{n-1})$ by $\zeta_n$, that is, ${\mathcal Z}_\rho (\theta; \phi_1,
\ldots,
\phi_{n-1}) = Z_-^{(n)} + \sin \theta \cdot I_{2^n} \otimes Z_+ + \cos
\theta \cdot \zeta_n (\phi_1, \ldots, \phi_{n-1})$.
In this case, $\zeta_n$ can be written using $\zeta_{n-1}$ as
\begin{align}
\zeta_{n} &= \cos \phi_{n-1} \cdot \sigma_3 \otimes \zeta_{n-1} \nonumber
\\ &+
\sin \phi_{n-1} \cdot i \sigma_2 \otimes ( I_{2^{n-1}} \otimes Z_+ ),
\label{zn+1}
\end{align}
with $\zeta_1 = i \sigma_2 \otimes Z_+$.

\end{description}

As is the case of $H_{16} (\theta, \phi)$, it is found that
${\mathcal Z}_\rho (\theta + d \theta; \phi_1,\ldots, \phi_{n-1})$ and
${\mathcal Z}_\rho (\theta; \phi_1,\ldots, \phi_{n-1})$
are not similar matrices if, at least, ${\rm Tr}
\, Z_+ \neq 0$ (if ${\rm Tr} \, Z_+ = 0$, it is rather difficult to determine
whether they are similar matrices or not).
Furthermore, it is found that ${\mathcal Z}_\rho (\theta; \phi_1',\ldots,
\phi_{n-1}')$ and
${\mathcal Z}_\rho (\theta; \phi_1,\ldots, \phi_{n-1})$ are similar matrices
as follows.

\begin{description}
\item[Proposition~2.] For $c_0'=c_0$, ${\mathcal Z} (c_0',c_1',\ldots,c_n')$
is obtained from the orthogonal transformation of
${\mathcal Z} (c_0,c_1,\ldots,c_n)$.

\item[Proof.]
For $n=1$, where
${\mathcal Z} (c_0,c_1) = \sigma_3 \otimes Z_- + (c_0 I_2 + c_1 i \sigma_2)
\otimes Z_+$, it is required that
$c_1'=c_1$ or $c_1'=-c_1$ under the condition of
$\sum_{i=0}^1 c_i^2 = \sum_{i=0}^1 c_i^{\prime 2} = 1$, with $c_0' = c_0$. In
the former case,
${\mathcal Z} (c_0,c_1)$ and ${\mathcal Z} (c_0,c_1')$ are identical.
In the latter case, we can write
$S {\mathcal Z} (c_0,c_1) S^{-1} = {\mathcal Z} (c_0,-c_1)$, with $S=
\sigma_3 \otimes I_{N}$. In this case, $S$ is an orthogonal matrix (${}^t \! S
= S^{-1}$).

For $n \geq 2$,
we will construct by induction the
generator
$T_{n-1}^{(i)} \; (= - {}^t  T_{n-1}^{(i)})$ such that
$\partial {\mathcal Z}_\rho / \partial \phi_i = [T_{n-1}^{(i)}, {\mathcal
Z}]$, namely
\begin{align}
\partial \zeta_n /
\partial
\phi_i &= [ T_{n-1}^{(i)}, \sec \theta \cdot Z_-^{(n)} +
\tan \theta \cdot
I_{2^n} \otimes Z_+ + \zeta_n]
%
%
\label{dz}
\end{align}
for $i=1, \ldots, n-1$.
Thus, it is sufficient to obtain $T_{n-1}^{(i)}$ such that
\begin{subequations}
\label{suff_cond}
\begin{align}
\label{suff1}
& \partial \zeta_n / \partial \phi_i = [T_{n-1}^{(i)}, \zeta_n] \\
\label{suff2}
& [T_{n-1}^{(i)}, Z_-^{(n)}] =
[T_{n-1}^{(i)}, I_{2^n} \otimes Z_+] = 0
\end{align}
\end{subequations}
for $i=1, \ldots, n-1$.
Substituting Eq.~(\ref{zn+1}) into Eq.~(\ref{suff1}), and comparing the
terms proportional to $\cos \phi_{n-1}$ and $\sin \phi_{n-1}$, we require
that
\begin{subequations}
\label{sub}
\begin{align}
& [T_{n-1}^{(i)}, \sigma_3 \otimes \zeta_{n-1} ]  \nonumber \\
&=
\begin{cases}
i \sigma_2 \otimes (I_{2^{n-1}} \otimes Z_+) \quad (\text{for } i = n-1),
\\
\sigma_3  \otimes \partial \zeta_{n-1} / \partial \phi_i
\quad
(\text{for } i=1,\ldots,n-2),
\end{cases}
\end{align}
and
\begin{align}
&[T_{n-1}^{(i)}, i \sigma_2 \otimes (I_{2^{n-1}} \otimes Z_+) ]   \nonumber
\\ &=
\begin{cases}
- \sigma_3 \otimes \zeta_{n-1} \quad (\text{for } i=n-1), \\
0 \quad (\text{for }i=1, \ldots,n-2),
\end{cases}
%
\label{sub1}
\end{align}
respectively.
On the other hand, Eq.~(\ref{suff2}) is rewritten as
\begin{align}
& [T_{n-1}^{(i)}, \sigma_3 \otimes Z_-^{(n-1)}]
=
[T_{n-1}^{(i)}, I_2 \otimes (I_{2^{n-1}} \otimes Z_+)] \nonumber \\
&= 0 \quad (\text{for } i=1, \ldots, n-1),
\label{sub2}
\end{align}
\end{subequations}
due to $Z_-^{(n)} = \sigma_3 \otimes Z_-^{(n-1)}$, and the associativity of
the direct product: $(A \otimes B ) \otimes C = A \otimes (B \otimes C)$.

Now we obtain $T_{n-1}^{(i)}$ inductively.
To begin with, we deal with the case of $n=2$, where $i=1$.
Suppose that $T_1^{(1)}$ can be decomposed as $T_1^{(1)} = \sigma_1 \otimes
\hat{T}_1$.
Then from Eq.~(\ref{sub}),
$\hat{T}_1$ should satisfy the following
relations:
\begin{align*}
& \{ \hat{T}_{1}, \, \zeta_{1}  \}  = - I_{2} \otimes Z_+, \\
& \{ \hat{T}_1, I_{2} \otimes Z_+ \} = \zeta_1,
\\
& \{ \hat{T}_1,  Z_-^{(1)} \} = 
[ \hat{T}_1, I_2 \otimes Z_+] = 0,
\end{align*}
where $\{ A, B \} = AB + BA$, and
use has been made of $[\sigma_i \otimes A, I_2 \otimes B] = \sigma_i \otimes
[A,B]$, and
$[\sigma_i \otimes A, \sigma_j \otimes B] = i \epsilon_{ijk} \sigma_k \otimes
\{ A, B \}$, with $e_{ijk}$ the totally antisymmetric tensor with
$\epsilon_{123} = 1$.
Noticing further that $\zeta_1 = i \sigma_2 \otimes Z_+$ and
$Z_-^{(1)} = \sigma_3 \otimes Z_-$, and using
$\{ \frac{1}{2} \sigma_2, \sigma_2 \} = I_2$,
$\{ \frac{1}{2} \sigma_2, I_2 \} = \sigma_2$,
$\{ \sigma_2, \sigma_3 \} = 0$, and
$[\sigma_2, I_2] = 0$,
we find that $\hat{T}_1$
can be chosen as $\hat{T}_1 = \frac{1}{2} i \sigma_2 \otimes I_N$. Thus, we
get
\begin{eqnarray*}
T_1^{(1)} = {\textstyle \frac{1}{2} } i
\sigma_1 \otimes \sigma_2
\otimes I_N.
\end{eqnarray*}
%
%
%

The next thing is to obtain $T_k^{(i)}$ ($i=1,\ldots,k$) under the condition
that there exists $T_{k-1}^{(i)}$ ($1=1, \ldots, k-1$) such that
Eq.~(\ref{suff_cond}) with $n=k$ holds, namely
\begin{subequations}
\label{n=k}
\begin{align}
%
& \partial \zeta_k / \partial \phi_i = [T_{k-1}^{(i)}, \zeta_k], \\
& [T_{k-1}^{(i)}, Z_-^{(k)}] =
[T_{k-1}^{(i)}, I_{2^k} \otimes Z_+] = 0
%
\end{align}
\end{subequations}
for $i=1, \ldots, k-1$.
On the other hand,
the equations for $T_k^{(i)}$ are given from Eq.~(\ref{sub}) by
\begin{subequations}
\label{n=k+1}
\begin{align}
\begin{split}
& [T_{k}^{(i)}, \sigma_3 \otimes \zeta_{k} ]  \\
&=
\begin{cases}
i \sigma_2 \otimes (I_{2^{k}} \otimes Z_+) \quad (\text{for }i = k),
 \\
\sigma_3  \otimes \partial \zeta_{k} / \partial \phi_i
\quad
(\text{for } i=1,\ldots,k-1),
\end{cases}
\end{split} \\
\begin{split}
& [T_{k}^{(i)}, i \sigma_2 \otimes (I_{2^{k}} \otimes Z_+) ]  \\
&=
\begin{cases}
- \sigma_3 \otimes \zeta_{k} \quad (\text{for } i=k),  \\
0 \quad (\text{for } i=1, \ldots,k-1),
\end{cases}
\end{split}
\\
\begin{split}
& [T_{k}^{(i)}, \sigma_3 \otimes Z_-^{(k)}] =
[T_{k}^{(i)}, I_2 \otimes (I_{2^{k}} \otimes Z_+)]  \\
&= 0 \quad \; (\text{for } i=1, \ldots, k).
\end{split}
\end{align}
\end{subequations}
Comparing Eqs.~(\ref{n=k}) and (\ref{n=k+1}), we find that $T_k^{(i)}$ for
$i=1, \ldots, k-1$ (not including $k$) can be written using $T_{k-1}^{(i)}$
as
\begin{align*}
T_k^{(i)} = I_2 \otimes T_{k-1}^{(i)} \quad (i=1,\ldots,k-1),
\end{align*}
due to $[I_2 \otimes A, \sigma_i \otimes B] = \sigma_i \otimes [A,B]$ and
$[I_2 \otimes A, I_2 \otimes B] = I_2 \otimes [A,B]$.
The existence of $T_k^{(i)}$ for $i=1,\ldots,k-1$ implies that ${\mathcal Z}
(c_0, \ldots, c_{k+1})$ and
${\mathcal Z} (c_0,\ldots,c_{k+1}) |_{\phi_1 = \ldots = \phi_{k-1} = 0}$
are similar matrices.
Thus,
the rest we have to do is to obtain $T_{k}^{(i)}$ for $i=k$, where we can
set
$\phi_1 =
\ldots = \phi_{k-1} = 0$ (because similarity
transformation is transitive).
Suppose that $T_k^{(k)}$ can be decomposed as $T_k^{(k)} = \sigma_1 \otimes
\hat{T}_k$, we obtain from Eq.~(\ref{n=k+1})
\begin{subequations}
\label{rest}
\begin{align}
& \{ \hat{T}_{k}, \, \zeta_{k}  \}  = - I_{2^k} \otimes Z_+,\\
& \{ \hat{T}_k, I_{2^k} \otimes Z_+ \} = \zeta_k,
\\
& \{ \hat{T}_k,  Z_-^{(k)} \} = 
[ \hat{T}_k, I_{2^k} \otimes Z_+] = 0.
\end{align}
\end{subequations}
Supposing further that
$\hat{T}_{k} = \sigma_3 \otimes \hat{T}_{k-1}$,
noticing that $\zeta_k = \sigma_3 \otimes
\zeta_{k-1}$ for
$\phi_1 =
\ldots = \phi_{k-1} = 0$, and using again
$Z_-^{(k)} = \sigma_3 \otimes Z_-^{(k-1)}$,
we find that $\hat{T}_{k-1}$ satisfies the same relations as
Eq.~(\ref{rest})
where
$k$ replaced by
$(k-1)$.
Repeating a
similar replacement, we can write $\hat{T}_k$ using $\hat{T}_1$, so
that $T_k^{(k)}$ can be written  as
\begin{align*}
T_k^{(k)} = {\textstyle \frac{1}{2} } i \sigma_1 \otimes
\underbrace{
\sigma_3 \otimes
\sigma_3 \otimes \ldots \otimes \sigma_3 \otimes}_{k-1} \sigma_2 \otimes I_N,
\end{align*}
for $\phi_1 = \ldots \phi_{n-1} = 0$.
Eventually, we have obtained $T_{n-1}^{(i)}$ ($i=1,\ldots,n-1$) for all $n \;
(\geq 2)$ inductively.

Since the generators $T_{n-1}^{(i)}$'s ($i=1,\ldots, n-1$) constructed here
are all antisymmetric, the corresponding similarity transformation turns out
to be an orthogonal one.
\hspace*{\fill} $\Box$

\end{description}

\subsection{Antisymmetricalness}

In this subsection, we give a reasonable condition to fix the parameter
$c_0$ in ${\mathcal Z} (c_0, \ldots, c_n)$ [or $\theta$ in $\mathcal Z_\rho
(\theta; \phi_1, \ldots, \phi_{n-1})$].
As far as ${\mathcal Z}$ $[={\mathcal Z} (c_0,\ldots,c_n)]$ and
${\mathcal Z}'$ $[={\mathcal Z} (c_0',
\ldots, c_n')]$ are not similar matrices, as is occurred for $c_0' = c_0 +
dc_0$,
some condition is necessary so that they may be equivalently represented.
Once $c_0$ is fixed, namely $c_0' = c_0$, we have the orthogonal
transformation
${\mathcal Z}' = S {\mathcal Z} {}^t \! S$.
%
%
%
%
%
%
This finding leads to the invariance of the following relation under
the orthogonal transformation $\mathcal Z \rightarrow S \mathcal Z
{}^t \! S$.

Consider the
transformation
${\mathfrak S} : {\mathbb R}^{4^n N^2}  \rightarrow
{\mathbb R}^m$ such that
${\mathfrak S} ({\mathcal Z})$ depends on $c_0$, namely
\begin{align}
{\mathfrak S} ({\mathcal Z}) = {\mathcal M} (c_0),
\label{sym}
\end{align}
where ${\mathcal M} (c_0)$ is a certain function of $c_0$.
If the limiting operator ${\mathfrak L}$ such that ${\mathfrak L}=
\lim_{ {\bf c}
\rightarrow {\bf c}'}$
[where ${\bf c}=(c_1,\ldots, c_n)$, ${\bf c}' = (c_1', \ldots, c_n')$, with
$| {\bf c} |^2 = | {\bf c}' |^2 = 1-c_0^2$] is commutative with the
transformation
${\mathfrak S}$, it is found that the relation of Eq.~(\ref{sym}) is
invariant under the orthogonal transformation
${\mathcal Z} \rightarrow S {\mathcal Z} {}^t
\! S$ as follows.
\begin{align*}
{\mathcal M} (c_0) &=
{\mathfrak L} ( {\mathcal M} (c_0) ) \\
&= {\mathfrak L} ( {\mathfrak S} ({\mathcal Z}) )  \\
&= {\mathfrak S} ( {\mathfrak L} ({\mathcal Z}) ) \\
&= {\mathfrak S} ( S {\mathcal Z} {}^t \! S ),
\end{align*}
where in the last line, use has been made of $S {\mathcal Z} {}^t \! S =
{\mathfrak L} ({\mathcal Z} )$.
However, Eq.~(\ref{sym}) does not necessarily remain invariant under any
orthogonal transformation, depending on the choice of ${\mathfrak S}$.
Thus, in fixing the parameter $c_0$, it seems reasonable that the condition
imposed on $\mathcal Z$ should be such that the relation of Eq.~(\ref{sym})
remains invariant under any orthogonal transformation.
That is, we impose the condition
\begin{align}
{\mathfrak S} ({\mathcal Z}) = {\mathfrak S} ( \Lambda {\mathcal Z} {}^t \!
\Lambda) \quad \text{for all } \Lambda \text{ such that } \Lambda {}^t \!
\Lambda = I,
\label{SZ}
\end{align}
to finally fix the parameter $c_0$.

Now we choose ${\mathfrak S}$ such that $[{\mathfrak S}, {\mathfrak L}] = 0$.
Due to the largeness of the number of the candidates for ${\mathfrak S}$, we
restrict ourselves to the case where ${\mathfrak S}$ is linear.
Furthermore, we concentrate ourselves on the cases of $m=1$ and $m=4^n N^2$. 
First, we deal with the case of $m=1$.
Considering that the diagonal elements of
${\mathcal Z}$ are all independent of
${\bf c}$, we find that
${\mathfrak S}$ can be chosen as ${\mathfrak S} = {\rm Tr}$, which
is linear and commutative with ${\mathfrak L}$. In this case, however,
Eq.~(\ref{SZ}) holds identically, so that the choice of
${\mathfrak S} = {\rm Tr}$ is not available in fixing $c_0$.
Next, we deal with the case of $m=4^n N^2$.
Noticing that
$M (c^{(n)} ) + {}^t \! M (c^{(n)} ) =
 M (c^{(n)} ) + M ( \overline{ c^{(n)} } ) =
 M ( c^{(n)} + \overline{ c^{(n)} } ) =
 M (2 c_0 e_0 )$,
we find that ${\mathfrak S}$ can be chosen as
\begin{align}
{\mathfrak S} ( {\mathcal Z} ) = {\mathcal Z} + {}^t \! {\mathcal Z} \;
(=2 c_0 I_{2^n} \otimes Z_+),
\label{choice}
\end{align}
which is linear and commutative with ${\mathfrak L}$.
In this case,
${\mathfrak S}$ is
commutative with
the orthogonal transformation,
namely,
${\mathfrak S} ( \Lambda {\mathcal Z} {}^t \! \Lambda) = \Lambda {\mathfrak
S} ( {\mathcal Z} ) {}^t \! \Lambda$,
so that
the condition of Eq.~(\ref{SZ}) is rewritten as
$[ {\mathfrak S} ( {\mathcal Z} ), \Lambda ] = 0$ for all $\Lambda$ such
that $\Lambda {}^t \! \Lambda = I$.
Rewriting $\Lambda = e^{\lambda A}$ ($\lambda \in \real$, $A=-{}^t
\! A$), we find that the above condition yields $[ {\mathfrak S}
({\mathcal Z}), A] = 0$ for all $A$ such that $A = -{}^t \! A$.
Furthermore,
putting $A=\left(
\begin{smallmatrix} 0 & A_1 \\ - {}^t \! A_1 & 0
\end{smallmatrix} \right)$, which is antisymmetric for all $A_1$, and
decomposing $\mathcal Z + {}^t \mathcal Z$ as $I_2
\otimes (2c_0 I_{2^{n-1}} \otimes Z_+)$, we obtain
%
$[2c_0 I_{2^{n-1}}
\otimes Z_+, A_1] = 0$ for all $A_1$. Thus, $2c_0 I_{2^{n-1}} \otimes Z_+$
should be 
proportional to the unit matrix, namely
\begin{eqnarray*}
2 c_0 Z_+ = k I_N,
\label{kI}
\end{eqnarray*}
where $k$ is a certain number.
For ${\mathcal Z}_+ \notin {\mathcal I}$ (where ${\mathcal I} = \{ x
I_N | x \in {\mathbb R } \}$), as in the case of the (original) Born-Infeld
Lagrangian, it follows that
$c_0$ (and hence, $k$)
is vanishing, so that ${\mathcal Z}$ turns out to be antisymmetric:
\begin{equation}
{\mathcal Z} + {}^t \! {\mathcal Z} = 0, \;\; c_0 = 0 \;\; (\mbox{for } Z_+
\not\in \; {\mathcal I}).
\label{anti}
\end{equation}

There remain $n$ parameters $(c_1, \ldots, c_n)$ such that $\sum_{i=1}^n c_i^2
= 1$.
However, it is found from Proposition~2, that ${\mathcal Z}
(0,c_1,\ldots,c_n)$ and
${\mathcal Z} (0,1,0,\ldots,0)$ are similar matrices, where
${\mathcal Z} (0,1,0,\ldots,0)$ is decomposed as
\begin{align}
&{} S^{-1} {\mathcal Z} (0, c_1, \ldots, c_n) S \nonumber \\
&=
{\mathcal Z} (0,1,0,\ldots, 0) \nonumber \\
&=
\underbrace{
{\mathcal Z} (0,1) \oplus {}^t \! {\mathcal Z} (0,1) \oplus \cdots \oplus
{}^{(t)} \! {\mathcal Z} (0,1)}_{2^n-1}.
\label{reduct}
\end{align}
For the Born-Infeld Lagrangian, Eq.~(\ref{reduct}) indicates the identity
\begin{equation}
| g + b F |^{2^n} =
\left| S
{\mathcal H} S^{-1} \right| \;\;\; (\mbox{for } n = 1,2,\ldots),
\label{final}
\end{equation}
where
\begin{eqnarray*}
{\mathcal H} = {\rm diag} \, (
\underbrace{
H_8 (0), - H_8 (0), \ldots, (\pm) H_8 (0)}_{2^{n-1}}
).
\label{H2n+2}
\end{eqnarray*}
Eventually, all the elements of ${\mathcal H}$ are given by those of $H_8
(0)$, despite the value of $n \; (\geq 1)$; no other variables are necessary
than the
8-dimensional phase space variables with non-commutative spacetime.

\subsection{Symmetricalness}

We have found that ${\mathcal Z}$ in Eq.~(\ref{mathZ}) turns out to be
antisymmetric under the condition such that is
invariant under any orthogonal transformation.
Even if the condition is neglected, ${\mathcal Z}$ cannot be symmetric unless
$Z_- = c_1 = \ldots = c_n =0$ (for the Born-Infeld Lagrangian, $Z_-=0$
corresponds to $F=0$, which is not physically allowed). However,
this does not necessarily mean that the determinant of a given $N$-dimensional
square matrix
$Z$ cannot be represented by the determinant of some
$2^n \cdot N$-dimensional symmetric matrix whose elements are given by the
linear combination of $Z_+$ and $Z_-$.
Consider, for example, the Born-Infeld Lagrangian ${\mathcal L}$.
Using the second identity in Eq.~(\ref{c7}), we can rewrite
${\mathcal  L}$ as
\begin{eqnarray*}
{\mathcal L} = | \tilde{H}_8 (\theta) |^{1/4}, \;\; \tilde{H}_8 (\theta) =
\sigma_3 \otimes g + U(\theta) \otimes bF,
%
\end{eqnarray*}
where $U(\theta) = \left( \begin{smallmatrix} \sin \theta & - \cos \theta \\
\cos \theta & \sin \theta \end{smallmatrix} \right)$, and
use has been made of
$| - (bF + g) | = |bF+g|$ due to the even dimensionality of $F$ and $g$.
Notice that $\tilde{H}_8 (\theta)$ is obtained by exchanging $g
\leftrightarrow bF$ in
$H_8 (\theta)$.
Apparently, $\tilde{H}_8 (0)$ is symmetric.
Considering that $H_8 (0)$ can be written as the commutator of the
8-dimensional phase space variables with non-commutative spacetime, one may
interpret the elements of $\tilde{H}_8 (0)$ as
the anticommutator of the corresponding variables
with ``non-anticommutative'' spacetime.

More generally, we have the following determinant identity.
\begin{description}
\item[Notation:]
$\tilde{\mathcal Z} = {\mathcal Z} (Z_+ \leftrightarrow Z_-)$.

\item[Corollary~4.] For $n \geq 1$,
\begin{eqnarray*}
(-1)^{N \cdot 2^{n-1}} | Z |^{2^n} = | \tilde{\mathcal Z} (c_0, \ldots, c_n)
|.
\end{eqnarray*}

\item[Proof.]
As mentioned in Remark~3, Eq.~(\ref{Zc0}) holds for arbitrary square matrices
$Z_+$ and $Z_-$. Thus, the following identity is obtained by replacing $Z_+$
($Z_-$) in Eq.~(\ref{Zc0}) by $Z_-$ ($Z_+$) as
\begin{align*}
| \tilde{\mathcal Z} (c_0, \ldots, c_n) | &=
\left( | Z_- + Z_+ | \left| {}^t \! Z_- - {}^t \! Z_+ \right|
\right)^{2^{n-1}} \\
&=
\left[
(-1)^N \, | Z_+ + Z_- |^2 \right]^{2^{n-1}} \\
&=
(-1)^{N \cdot 2^{n-1} } | Z |^{2^n},
\end{align*}
which is to be proven. \hspace*{\fill} $\Box$

\item[Remark~4:]
For $N$ even, $(-1)^{N \cdot 2^{n-1}}$ turns out to be unity. In this case,
$\tilde{\mathcal Z}$ can be regarded as another solution for ${\mathcal Z}$.

\end{description}

Different from ${\mathcal Z}_\rho (\theta; \phi_1, \ldots, \phi_{n-1})$, it
is rather difficult to show that $\tilde{\mathcal Z}_\rho (\theta + d \theta;
\phi_1, \ldots, \phi_{n-1})$ and
$\tilde{\mathcal Z}_\rho (\theta; \phi_1, \ldots, \phi_{n-1})$ are not
similar matrices, because we cannot derive the contradiction from $\partial
\tilde{\mathcal Z}_\rho / \partial \theta = [\tilde{T}_\theta,
\tilde{\mathcal Z}_\rho]$ by taking the trace in both sides, due to the
tracelessness of an antisymmetric matrix
$Z_-$.
(For the Born-Infeld Lagrangian, for example, it is found from direct
calculation, that
$\tilde{H}_8 (0)$ and
$\tilde{H}_8 (d \theta)$ are not similar matrices, in spite of the relation of
${\rm Tr} \, [ \partial \tilde{H}_8 (0) / \partial \theta ] = 0$.) Despite the
difficulty, as far as two $\tilde{\mathcal Z}$'s are not similar matrices, we
should impose some condition so that they turn out to be equivalently
represented. In so doing, we will follow an analogous procedure discussed in
the previous subsection.
All the relations concerning $\tilde{\mathcal Z}$ can be obtained by
replacing ${\mathcal Z}$ and ${}^t \! {\mathcal Z}$ in the relations in the
previous subsection by
\begin{eqnarray*}
({\mathcal Z}, {}^t \! {\mathcal Z}) \rightarrow (\tilde{\mathcal Z},
- {}^t \! \tilde{\mathcal Z}),
\end{eqnarray*}
because under the substitution of
$Z_+ \leftrightarrow Z_-$, where
${\mathcal Z}$ transforms to $\tilde{\mathcal Z}$,
${}^t \! {\mathcal Z}$ transforms to $- {}^t \! \tilde{\mathcal Z}$. Thus,
the condition corresponding to Eq.~(\ref{anti}) leads to
\begin{eqnarray*}
\tilde{\mathcal Z} - {}^t \! \tilde{\mathcal Z} = 0, \;\; c_0 = 0,
\end{eqnarray*}
that is, the symmetricalness of $\tilde{\mathcal Z}$ is obtained under the
condition that the relation of $\mathfrak S (\tilde{\mathcal
Z}) = \tilde{\mathcal Z} - {}^t
\!
\tilde{\mathcal Z}$ remains invariant under any
orthogonal transformation $\tilde{\mathcal Z} \rightarrow \Lambda
\tilde{\mathcal Z}
\, {}^t \! \Lambda$.

Once $\tilde{\mathcal Z} (c_0, \ldots, c_n)$ is chosen as symmetric, where
$c_0$ is required to be vanishing, it is found from Proposition~2 that
$\tilde{\mathcal Z} (0,c_1,\ldots,c_n)$ and $\tilde{\mathcal Z} (0,c_1',
\ldots, c_n')$ are related through the same orthogonal transformation as for
${\mathcal Z} (0,c_1,\ldots,c_n)$ and ${\mathcal Z} (0,c_1',\ldots,c_n')$,
because the generator
$T_k^{(i)}$ can be chosen as the same, despite the
substitution of
$Z_+ \leftrightarrow Z_-$.


\section{Summary}

We have rewritten the Born-Infeld Lagrangian, which is originally given by
the determinant of a $4 \times 4$ matrix composed of the metric
tensor
$g$ and the field strength tensor $F$, using the determinant of a
$8 \times 8$ matrix $H_8$.
If the elements of $H_8$ are
given by the linear combination of $g$ and $F$ (this means that the elements
are rank-2 tensors of the same type as $g$ and $F$),
it is found, based on the representation matrix for the right-multiplication
operator of the Cayley-Dickson algebras, that
$H_8$ has a single parameter $\theta$, where $H_8 (\theta + d\theta)$ is not
equivalent to
$H_8 (\theta)$ in the sense that they cannot be obtained from each other
through a similarity transformation.

The determinant identity is generalized to
a $4 \cdot 2^n \times 4 \cdot 2^n$ matrix
$H_{4 \cdot 2^n}$ (Proposition~1). As $n$ increases, the number of
the parameters in $c^{(n)}$ of $H_{4 \cdot 2^n}$ increases one by one.
However, all the parameters except one can be fixed due to the orthogonal
transformation for $H_{4 \cdot 2^n}$ (Proposition~2).
In the orthogonal transformation,
the spinor
representation is realized, which is attributed to the associativity
of $c^{(n)}$ (Remark~2).
Under a reasonable condition to fix the single parameter,
$H_{4\cdot 2^n}$ can be chosen as antisymmetric (or symmetric, by exchanging
$g$ and $F$).
Once $H_{4 \cdot 2^n}$ is chosen as antisymmetric, all the elements of $H_{4
\cdot 2^n}$ can be given by the orthogonal transformation of $H_8 (0)$.
In this sense, no variables are necessary other than the elements of $H_8
(0)$.


\section*{Aacknowledgments}

The author is indebted to Prof. R. Takahashi for his useful discussion and
suggestion.



\appendix

\section{$| M (x) |, \; x \in \alg_4$}

%
%
For $x = (x_1, x_2)$ in $\alg_4 = \alg_{3} \times \alg_{3}$,
it is found from Lemma~5 that
\begin{align}
| M (x) | &= | k I_8 + i B (x_1,x_2) | \nonumber \\
&= k^8 \left| I_8 + i B (x_1/ \sqrt{k}, x_2 / \sqrt{k}) \right|,
\label{c8}
\end{align}
where $k= \| x \|^2 = \| x_1 \|^2 + \| x_2 \|^2$.
Putting $f (x_1) = (x_1^0, \ldots, x_1^7)$ and
$f (x_2) = (x_2^0, \ldots, x_2^7)$, we can write $B (x_1,x_2)$ as
%
%
\begin{eqnarray}
\frac{1}{2} B (x_1,x_2)
=
\left(
\begin{array}{cccccccc}
0 & 0 & 0 & 0 & 0 & 0 & 0 & 0\\
  & 0 & [4,7] + [6,5] & [6,4] + [7,5] & [3,6]+[7,2] & [2,6]+[3,7]
& [4,3] + [5,2] & [2,4] + [5,3] \\
 & & 0 & [4,5]+[7,6] & [1,7]+[5,3] & [3,4]+[6,1] & [1,5]+[3,7]
& [4,1] + [6,3] \\
& & & 0 & [2,5]+[6,1] & [4,2]+[7,1] & [1,4]+[7,2] & [1,5]+[2,6] \\
& & & & 0 & [2,3] + [7,6] & [3,1] + [5,7] & [1,2] + [6,5] \\
& & & \mbox{($\ast)$} & & 0 & [2,1] + [7,4] & [3,1] + [4,6] \\
& & & & & & 0 & [3,2] + [5,4] \\
& & & & & & & 0
\end{array}
\right),
\label{1/2B}
\end{eqnarray}
%
%
%
where $[n,m] = x_1^n x_2^m - x_1^m x_2^n$, and $(\ast)$ can be obtained from
${}^t \! B (x_1,x_2) = - B(x_1,x_2)$.
Substituting Eq.~(\ref{1/2B}) into (\ref{c8}), we obtain
\begin{equation}
| M (x) | = k^4 (k^2 - \mu^2)^2,
\label{a3}
\end{equation}
where 
\begin{align}
\mu^2 &= 4 \sum_{1 \leq n < m \leq 7} [n,m]^2 \nonumber \\
&=
4 \left[ | {\bf x}_1 |^2 | {\bf x}_2 |^2 - ({\bf x}_1 \cdot {\bf x}_2)^2
\right],
\label{a4}
\end{align}
with ${\bf x}_1 = (x_1^1, \ldots, x_1^7)$ and
${\bf x}_2 = (x_2^1, \ldots, x_2^7)$.

From Eq.~(\ref{a3}), it is implied that there exists $x \; (\neq 0)$ such
that $| M (x) | = 0$, namely, a zero divisor.
Actually, the zero divisor is realized if and only if
$| M (x) | = 0$ (we do not distinguish the right and left zero divisors due
to $|M (x)| = | \bar{M} (x)|$).
Using Eq.~(\ref{a4}), we can rewrite $k^2 - \mu^2$ as
\begin{align*}
k^2 - \mu^2 &= r^4 + 2r^2 (|{\bf x}_1 |^2 + | {\bf x}_2 |^2) \nonumber \\
&+
(|{\bf x}_1 |^2 - | {\bf x}_2 |^2 )^2 + 4 ({\bf x}_1 \cdot {\bf x}_2)^2,
\end{align*}
where $r^2 = (x_1^0)^2 + (x_2^0)^2$.
Thus, the zero divisor for $x \in \alg_4$ is realized for $k^2 - \mu^2 = 0$
(with $x \neq 0$), that is
\begin{eqnarray*}
x_1^0 = x_2^0 = | {\bf x}_1 | - | {\bf x}_2 | = {\bf x}_1 \cdot {\bf x}_2 = 0,
\; \mbox{with }|{\bf x}_1 | \neq 0.
\end{eqnarray*}
The condition of $x_1^0 = x_2^0 = 0$, which is sometimes called doubly pure
(imaginary), is one of the characteristics of the zero divisor for $x \in
\alg_n$ ($n \geq 4)$~\cite{moreno}.

For $x \in \alg_n$ ($n \geq 5$), $|M (x)|$ is
more and more complicated to calculate, because not only $B(x_1, x_2)$ but
also $A(x_1,x_2)$ is not, in general, diagonal.

\section{Determinant identities}

For square matrices $A, B$, we obtain

\begin{align}
\left|
\begin{array}{cc}
A & -B\\
B & -A
\end{array}
\right| &=
\left|
\begin{array}{cc}
A+B & -B\\
B+A & -A
\end{array}
\right|  \nonumber \\
&=
\left|
\begin{array}{cc}
A+B & -B\\
0 & -A + B
\end{array}
\right| \nonumber \\
&=
| A + B | \, |B-A| \nonumber \\
&=
| A + B | \, |{}^t \! B - {}^t \! A |.
\end{align}
In a similar way,
\begin{eqnarray}
\left|
\begin{array}{cc}
A & B\\
-B & A
\end{array}
\right| = | A + i B | | {}^t \! A -i {}^t \! B |.
\label{b2}
\end{eqnarray}
More generally, for $n \times n$ matrices
$A,B,C,D$
\begin{align}
\left|
\begin{array}{cc}
A & B\\
C & D
\end{array}
\right| &= | D | \, |A - B D^{-1} C | \nonumber \\
&=
| - B | \, | C - DB^{-1} A |,
\label{c7}
\end{align}
which is derived from
$\left(
\begin{smallmatrix}
A & B\\ C & D
\end{smallmatrix}
\right) = \left(
\begin{smallmatrix}
I & B\\ 0 & D
\end{smallmatrix}
\right)
\left(
\begin{smallmatrix}
A - BD^{-1}C & 0\\ D^{-1} C & I
\end{smallmatrix}
\right)$, and
$\left(
\begin{smallmatrix}
A & B\\ C & D
\end{smallmatrix}
\right) =
(-1)^n
\left(
\begin{smallmatrix}
C & D\\ A & B
\end{smallmatrix}
\right)$.

\section{Derivation of Eq.~(\ref{F})}

From Lemma~2 and the identity $M(x) = \bar{M} (x) = x^0$ for $x = x^0 e_0$
in $\alg_0$, we obtain $M (x) = \bar{M} (x) = \left(
\begin{smallmatrix} x^0 & x^1 \\ - x^1 & x^0
\end{smallmatrix} \right)$ for $x = x^0 e_0 + x^1 e_1$ in $\alg_1$.
Similarly, we obtain
%
%
\begin{align}
M (x) = \left(
\begin{array}{cccc}
x^0 & x^1 & x^2 & x^3\\
-x^1 & x^0 & -x^3 & x^2\\
-x^2 & x^3 & x^0 & -x^1\\
-x^3 & -x^2 & x^1 & x^0
\end{array} \right), \quad
\bar{M} (x) = \left(
\begin{array}{cccc}
x^0 & x^1 & x^2 & x^3\\
-x^1 & x^0 & x^3 & -x^2\\
-x^2 & -x^3 & x^0 & x^1\\
-x^3 & x^2 & -x^1 & x^0
\end{array} \right)
\label{mx4}
\end{align}
%
for $x = \sum_{i=0}^3 e^i e_i$ in $\alg_2$,
so that
\begin{align}
{\rm Ad} (x) &\equiv \bar{M} (x) - M (x) \nonumber \\
&= 2
\left(
\begin{array}{cccc}
0 & 0 & 0 & 0\\
0 & 0 & x^3 & -x^2\\
0 & -x^3 & 0 & x^1\\
0 & x^2 & -x^1 & 0
\end{array} \right).
\label{ad}
\end{align}
Analogous representation matrices for $\alg_2$ are found in
Ref.~\cite{yano}.
On the other hand, $F_{\mu \nu}$ and $F^{\mu \nu}$ are given [with the
signature of the Minkowski metric $(+,-,-,-)$] by
%
%
\begin{align}
F_{\mu \nu} = \left(
\begin{array}{cccc}
0 & E_x & E_y & E_z\\
-E_x & 0 & -B_z & B_y\\
-E_y & B_z & 0 & -B_x\\
-E_z & -B_y & B_x & 0
\end{array} \right), \quad
F^{\mu \nu} = \left(
\begin{array}{cccc}
0 & -E_x & -E_y & -E_z\\
E_x & 0 & -B_z & B_y\\
E_y & B_z & 0 & -B_x\\
E_z & -B_y & B_x & 0
\end{array} \right).
\label{FF}
\end{align}
%
%
%
Comparing Eq.~(\ref{FF}) with Eqs.~(\ref{mx4}) and (\ref{ad}), we
obtain Eq.~(\ref{F}).


\end{document}